\documentclass[english,twocolumn,showpacs,prb]{revtex4}

\usepackage[T1]{fontenc}
\usepackage[latin9]{inputenc}
\usepackage{graphicx}

\usepackage{color}

\makeatletter
\@ifundefined{textcolor}{}
{%
 \definecolor{BLACK}{gray}{0}
 \definecolor{WHITE}{gray}{1}
 \definecolor{RED}{rgb}{1,0,0}
 \definecolor{GREEN}{rgb}{0,1,0}
 \definecolor{BLUE}{rgb}{0,0,1}
 \definecolor{CYAN}{cmyk}{1,0,0,0}
 \definecolor{MAGENTA}{cmyk}{0,1,0,0}
 \definecolor{YELLOW}{cmyk}{0,0,1,0}
 }

\makeatother

\usepackage{babel}

\begin{document}

\title{The dependence of topological Anderson insulator on the type of disorder}
%\author{Juntao Song$^{1}$, Hua Jiang$^{2}$, Haiwen Liu$^{3}$, Qing-feng Sun$^{3}$  and X. C. Xie$^{2}$}
\author{Juntao Song$^{1}$, Haiwen Liu$^{2}$, Hua Jiang$^{3,\ast}$,  Qing-feng Sun$^{2}$  and X. C. Xie$^{3}$}
\affiliation{$^1$Department of Physics and Hebei Advanced Thin Film  Laboratory, Hebei Normal University, Hebei
050024, China \\
$^2$Beijing National Lab for Condensed Matter Physics and Institute
of Physics, Chinese Academy of Sciences, Beijing 100080, China \\
$^3$International Center for Quantum Materials, Peking University, Beijing
10087, China}
}

\date{\today}

\begin{abstract}
This paper details the investigation of the influence of different disorders in two-dimensional topological insulator systems. Unlike the phase transitions to topological Anderson insulator induced by normal Anderson disorder, a different physical picture arises when bond disorder is considered. Using Born approximation theory, an explanation is given as to why bond disorder plays a different role in phase transition than does Anderson disorder. By comparing phase diagrams, conductance, conductance fluctuations, and the localization length for systems with different types of disorder, a consistent conclusion is obtained. The results indicate that a topological Anderson insulator is dependent on the type of disorder. These results are important for the doping processes used in preparation of topological insulators.
\end{abstract}


\maketitle

\section{Introduction}
With research breakthroughs in HgTe/CdTe quantum wells and 3D topological materials,~\cite{1,2,3,4,5,6,7}  topological insulators have attracted much attention in recent years. These unique topological properties are responsible for some interesting and surprising phenomena. For example, in 2009, Li, Chu, Jain and Shen~\cite{Shen} discovered that Anderson disorder can lead to a topological phase transitions with quantized conductance and named this phase topological Anderson insulator (TAI). The phenomenon of TAI was also reported by Jiang {\sl et al.}~\cite{Jiang} in their work. Subsequently, the origins of TAI, as well as TAI in other systems, have been studied by many groups using a variety of methods.~\cite{Guo,Xing,Kuramoto,Beenakker,Prodan,Chen,Zhang}

Guo {\sl et al.}~\cite{Guo}  found that the TAI phenomenon also exists in disordered 3D topological insulator. Recently, Xing {\sl et al.}~\cite{Xing} compared the disorder effects in three different systems where the quantum anomalous Hall effect exists. They observed TAI in these three different systems and demonstrated that increasing disorder strength produces the TAI phenomenon. Furthermore, Yamakage  {\sl et al.}~\cite{Kuramoto} reported similar phase transitions in disordered $Z_2$ topological insulators when considering $s_z$ non-conserving spin-orbit coupling.

The origins of TAI have been studied by many groups as well. After the initial reports of TAI,~\cite{Shen} Jiang {\sl et al.}~\cite{Jiang} calculated the distribution of local currents in real space at various strengthes of Anderson disorder and provided an explanation on TAI phase by studying the helical edge states in HgTe/CdTe quantum wells. In addition, it has been shown by Groth {\sl et al.}~\cite{Beenakker} that the phase transition to TAI should ascribe to a negative correction to the
topological mass because of Anderson disorder. Using the effective medium theory and through numerical calculations, Groth {\sl et al.}~\cite{Beenakker} discussed in detail how Anderson disorder renormalizes the topological mass, chemical potential and finally induces a phase transition. This interpretation provides a clear physical image of TAI and has been accepted generally by most physicists in this field.

The TAI effect has also been identified in 3D topological insulators and the effective medium theory noted above is found to be useful when describing the 3D case as reported in the paper by Guo {\sl et al.}~\cite{Guo} However, by looking at the phase diagrams, Prodan~\cite{Prodan} takes the position in his paper that TAI should not be considered a
distinct phase but should be described as part of the quantum spin-Hall phase. In addition, Chen {\sl et al.}~\cite{Chen} discussed the TAI phenomenon from the point of band structures. Finally, by calculating the $Z_2$ topological number in systems with periodic disordered supercell regimes, Zhang {\sl et al.}~\cite{Zhang} verified that TAI corresponds to a topologically non-trivial phase.

Thus, it has been well documented that Anderson disorder may induce TAI. However, it is not clear whether TAI is certain to be observed experimentally and whether all types of disorder can definitely push a phase transition to TAI for an anomalous quantum Hall system. It is the purpose of this paper to address these questions.

Besides the on-site Anderson disorder, there exists another type of disorder, bond disorder, which originates from the deformation of the lattice or from some other interactions that induce a random hopping term. This type of disorder exists widely and cannot be ignored when describing a real system. This is especially the case when considering a 2D system such as graphene. We note that such bond disorders have been studied extensively in various systems.~\cite{Jia,Altland,Ostrovsky,Nagaosa} Following the effective medium theory,~\cite{Beenakker} it is shown that, unlike Anderson disorder which appears through $\sigma_0$ or $\sigma_z$ term in Hamiltonian, bond disorder, which appears through $\sigma_x$ or $\sigma_y$ term
in Hamiltonian, would renormalize the topological mass by adding a positive mass correction. Therefore, according to the effective medium theory, TAI phenomenon cannot arise. To further investigate this issue, we will describe two different models. One is the HgTe/CdTe quantum wells and the other is the Haldane model.~\cite{Haldane} These models have been chosen because the first model is always used to study the TAI phenomena, and in the second model the $\sigma_x(\sigma_y)$ disorder can be easily introduced by the deformation of a honeycomb lattice and the $\sigma_x(\sigma_y)$ disorder comes very naturally if there is a random hopping correction to the nearest hopping term. We have investigated the phase diagram, conductance, conductance fluctuations and the localization length for these two concrete models. It will be shown that all of the results are consistent and that $\sigma_x(\sigma_y)$ disorder cannot lead to TAI and this type of disorder prohibits the TAI phenomenon in some sense. Therefore, in a real system, the presence of TAI phenomena may be determined by which type of the $\sigma_x(\sigma_y)$ and $\sigma_0(\sigma_z)$ disorder is stronger.

The rest of this paper is organized as follows: In Sec.~\ref{sec:models}, we introduce the two models in the tight-binding representation and derive the formulas of the conductance, the renormalized topological mass $\overline{M}$ and the renormalized chemical potential $\overline{\mu}$. The numerical results are discussed in Sec.~\ref{sec:discussions}. Finally, a brief summary is given in Sec.~\ref{sec:conclusions}.

\section{Theoretical Models}
\label{sec:models}
The first model, which has been studied extensively, is the standard HgTe/CdTe quantum wells.
The tight-binding Hamiltonian for the square lattice sketched in Fig.1(a) has the form:~\cite{Jiang}
\begin{eqnarray}
H_1 &=&\sum_{\mathbf{i}}\varphi _{\mathbf{i}}^{\dagger }\left(
\begin{array}{cccc}
E_s & U_i & 0 & 0 \\
U_i & E_p & 0 & 0 \\
0 & 0 & E_s & U_i \\
0 & 0 & U_i & E_p \\
\end{array}
\right) \varphi _{\mathbf{i}}  \nonumber \\
&+&\sum_{\mathbf{i}}\varphi _{\mathbf{i}}^{\dagger }\left(
\begin{array}{cccc}
V_{ss} & V_{sp}^{\ast } & 0 & 0 \\
-V_{sp} & V_{pp} & 0 & 0 \\
0 & 0 & V_{ss} & V_{sp} \\
0 & 0 & -V_{sp}^{\ast } & V_{pp} \\
\end{array}
\right) \varphi _{\mathbf{i}+\delta x}+h.c.  \nonumber \\
&+&\sum_{\mathbf{i}}\varphi _{\mathbf{i}}^{\dagger }\left(
\begin{array}{cccc}
V_{ss} & iV_{sp} & 0 & 0 \\
 iV_{sp}^{\ast } & V_{pp} & 0 & 0 \\
0 & 0 & V_{ss} & -iV_{sp}^{\ast } \\
0 & 0 & -iV_{sp} & V_{pp}\\
\end{array}
\right) \varphi _{\mathbf{i}+\delta y}+h.c.  \nonumber \\
\end{eqnarray}
Here $\mathbf{i}=(\mathbf{i}_x,\mathbf{i}_y)$ is the site index, and $\delta x$ and $
\delta y$ are unit vectors along the $x$ and $y$ directions.
$\varphi _{
\mathbf{i}}=(a_{\mathbf{i}},c_{\mathbf{i}},b_{\mathbf{i}},d_{\mathbf{i}
})^{T} $ represents the four annihilation operators of the electron on
the site $ \mathbf{i}$ with the state indices $|s,\uparrow \rangle
$,$ |p_{x}+ip_{y},\uparrow \rangle $ , $|s,\downarrow \rangle $, $
|-(p_{x}-ip_{y}),\downarrow >$ respectively. $E_{s},E_{p},V_{ss},V_{pp},$ and $V_{sp}$
are the five independent parameters that characterize the clean
HgTe/CdTe samples. $U_{i}$ represents random bond disorder, which is uniformly
distributed in the range $[-\frac{U}{2},\frac{U}{2}]$ with the disorder strength $U$.~\cite{Jia,Altland,Ostrovsky,Nagaosa}
Note that, in real materials, the disorder strength in same sites
should be much stronger than that between neighbor sites and therefore only random bond disorder for the same cell is included.~\cite{Note21}
It is clear that near the $\mathbf{\Gamma} $ point, the lattice
Hamiltonian [Eq. (1)] in $\mathbf{k}$-representation can be reduced to
the continuous Hamiltonian in Ref.6 when we take $V_{sp}=-iA/2a$, $
V_{ss}=(B+D)/a^{2}$, $V_{pp}=(D-B)/a^{2}$, $E_{s}=C+M-4(B+D)/a^{2}$,
and $ E_{p}=C-M-4(D-B)/a^{2}$.
\begin{figure}[!ht]
\includegraphics[width=1.0\columnwidth]{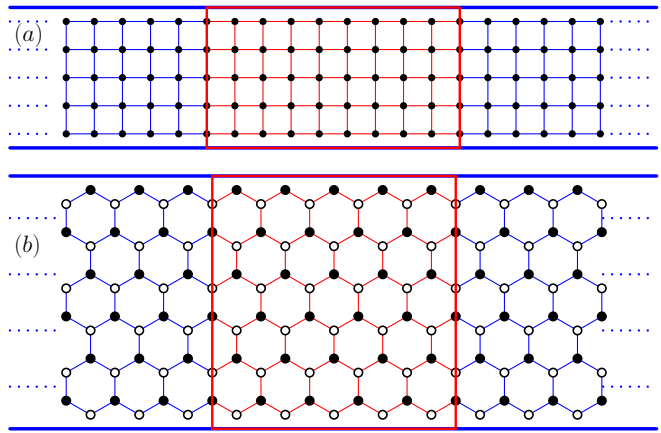}
\caption{
(Color online) A schematic diagram of a infinitely long
ribbon on the square lattice for the HgTe/CdTe model (a) and a infinitely
long ribbon in the honeycomb lattice along the zigzag direction for the
Haldane model (b).}
\end{figure}

Here $a$ is the lattice constant and all of the parameters $A$, $B$, $C$, $D$, and $M$ can be controlled
experimentally\cite{5}. The topological mass $M$ can be tuned continuously
by changing the thickness of the HgTe and subsequently switches the HgTe/CdTe wells between a
topologically nontrivial phase and a topologically trivial phase.
In this model, the individual spin-up Hamiltonian
and spin-down Hamiltonian in Eq. (1) are time-reversal
symmetric to each other. Since they are decoupled, we can deal
with them individually. For simplicity, we shall focus only on the spin-up
Hamiltonian in the following calculations.

The Haldane model proposed by Haldane in 1988,~\cite{Haldane} considers honeycomb lattice
with next-nearest-neighbor coupling and a staggered sublattice potential. The Hamiltonian
can be expressed as:~\cite{Haldane}
\begin{eqnarray}
H_2 & =&\sum_{m}{\varepsilon_m c^\dagger_m c_m}+\sum_{\langle m,n\rangle}
{t_{mn}c^\dagger_{m}c_{n}}+t_2\sum_{\langle\langle m,n\rangle\rangle}
e^{i\upsilon_{mn}\phi}{c^\dagger_{m}c_{n}}.\nonumber\\
\end{eqnarray}
The first term is the onsite energy and $\varepsilon_m=\pm \Delta$ for the A site (B site),
which is shown by a dot (circle) in Fig. 1(b).
The second term is the usual nearest neighbor hopping term. Here,
random bond disorder is introduced by
$t_{mn}=t_{nm}=t_1+ \delta t_{r}$, where $\delta t_{r}$ is uniformly
distributed in the range $[-\frac{U}{2},\frac{U}{2}]$ with
the disorder strength $U$. Note that the random hopping term gives the
bond disorder a more concrete physical image of the underlying nature of the term.
The third term is the second neighbor hopping
term with bond dependent phase. Note that $\upsilon_{mn}$ is different for different hopping directions.~\cite{1,Haldane,Xing} For example, if an electron in A or B site makes a left (right) turn to get to the second site, $\upsilon_{mn}=1 (-1)$.

By performing Fourier transformations, we can easily obtain the low energy effective
Hamiltonian for the two models given above. Note that the low energy effective Hamiltonian for the
two models have the same representation and can be written as follows:~\cite{6}
\begin{eqnarray}
H(\hat{k}) & =&H_0(\hat{k})+V=\alpha(\hat{k}_x \sigma_x-\hat{k}_y \sigma_y)\nonumber\\
&+&(M+\beta \hat{k}^2)\sigma_z+(\mu+\gamma \hat{k}^2)\sigma_0+U(r)\sigma_d,
\end{eqnarray}
where the operator $\hat{k}_{x/y}$ can be represented with the
momentum operator: $\hat{k}_{x/y}=-i\hbar \partial_{x/y}$.
This Hamiltonian is a two-dimensional Dirac Hamiltonian where the three Pauli matrices $\sigma_x$, $\sigma_y$, $\sigma_z$ and the unit matrix $\sigma_0$ represent the pseudospins. Note that the pseudospin is formed by the $s$ or $p$ orbital (corresponding to the A or B sublattice) for the HgTe/CdTe (Haldane) model. The scalar potential $U(r)$ accounts for the disorder amplitude and $\sigma_d=\sigma_{0/x}$ denotes Anderson disorder or bond disorder respectively. The parameters $\alpha$, $\mu$, $\gamma$ and $\beta$ for the HgTe/CdTe model have the simple form of:
\begin{eqnarray}
\alpha=-A;\hspace{2mm} \mu=C;\hspace{2mm} \gamma=-D; \hspace{2mm}
\beta=-B.
\end{eqnarray}

Meanwhile, the low energy effective Hamiltonian for the Haldane model can be expanded at the two inequivalent Dirac points $K$ and $K^*$. Therefore, the parameters $\alpha$, $\mu$, $M$, $\gamma$ and $\beta$ for the Haldane model can be represented as:
\begin{eqnarray}
\alpha&=&\frac{3}{2}t_1a;\hspace{2mm} \mu=-\sqrt{3}t_2cos\phi; \hspace{2mm} M=\Delta\pm3\sqrt{3}t_2sin\phi,\nonumber\\
\gamma&=&\frac{9}{4}t_2a^2cos\phi;\hspace{2mm}
\beta=\mp\frac{9\sqrt{3}}{4}t_2a^2sin\phi,
\end{eqnarray}
where $\pm$ or $\mp$ corresponds to $K$ or $K^*$ respectively.

It has been explained~\cite{Beenakker} that elastic scattering by onsite Anderson disorder
(called $\sigma_0$ disorder to distinguish it from bond disorder being called $\sigma_{x}$
disorder correspondingly in this paper)
 causes a state with a definite momentum to decay exponentially as a function of space  and time.
 Therefore, a negative correction to the topological mass $M$ is induced due to the quadratic term
 in the Hamiltonian. The renormalized mass $\overline{M}=M+\delta M$ may, in fact, have the opposite sign to
 the bare mass $M$. This implies that onsite Anderson disorder, $\sigma_0$, may induce
 a phase transition from a normal insulator to the topological phase. When this happens, TAI phenomena can be
 observed. However, the question arises as to whether this applies to the bond disorder, namely $\sigma_x$ disorder. We will now
 qualitatively show this mechanism through the following derivation.

First, we begin with the Hamiltonian in Eq. (3). Following the same derivation given in the paper by Groth {\sl et al.},~\cite{Beenakker}
the self-energy can be obtained from the equation:
\begin{eqnarray}
[E-H_0(k)-\Sigma(E,k)]^{-1} & =& \langle [E-H(k)]^{-1} \rangle,
\end{eqnarray}
where $\langle\ldots\rangle$ denotes the disorder average. Then, the self-energy can be represented in
the form: $\Sigma(E,k)=\Sigma_0 \sigma_0+\Sigma_x \sigma_x+\Sigma_y \sigma_y+\Sigma_z \sigma_z$.
Obviously,  the role of $\Sigma(E,k)$ induced by $\sigma_0$ or $\sigma_x$ disorder must be to
make a correction to the $\sigma_i$ terms in $H(k)$. The topological mass and the chemical potential are then renormalized and take the form:
\begin{eqnarray}
\overline{M}=M+\lim_{k\rightarrow0}Re\Sigma_z,\ \ \ \ \
\overline{\mu}=E-\lim_{k\rightarrow0}Re\Sigma_0.
\end{eqnarray}
To acquire self-energy $\Sigma(E)$ in numerical calculations, the self-consistent Born approximation
 is applicable and the integral equation for $\Sigma(E)$ and can be written as:
\begin{eqnarray}
\Sigma(E)
=\frac{U^2}{12}(a/2\pi)^2\int_{BZ} d^2k [\sigma_d(E^+I-H_0-\Sigma)^{-1}\sigma_d],
\end{eqnarray}
where $\sigma_d=\sigma_{0/x}$ denotes the type of disorder.

Up this point, it becomes apparent that the different types of disorder may lead to
a different correction to
the topological mass, the sign of which is critical for classifying
 the topological phase of the system. To see this clearly, we neglect $\Sigma(E)$ on the
 right side of Eq. (8).  An approximate solution for $\Sigma(E)$ with a closed form is then given by:
\begin{eqnarray}
\overline{M}=M\mp\frac{U^2a^2}{48\pi\hbar^2}\frac{\beta}{\beta^2-\gamma^2}
ln\mathbf{\bigg{|}}\frac{\beta^2-\gamma^2}{E^2-M^2}\mathbf{\bigg{|}}
(\frac{\pi\hbar}{a})^4\\
\overline{u}=E-\frac{U^2a^2}{48\pi\hbar^2}\frac{\gamma}{\beta^2-\gamma^2}
ln\mathbf{\bigg{|}}\frac{\beta^2-\gamma^2}{E^2-M^2}\mathbf{\bigg{|}}
(\frac{\pi\hbar}{a})^4\\ \nonumber
\end{eqnarray}
where $\mp$ corresponds to $\sigma_0$ and $\sigma_x$ disorder
respectively. For random bond disorder, the $\sigma_0$ term in
the self-energy $\Sigma(E,k)$ above is obtained by
$\sigma_x\sigma_0\sigma_x=\sigma_0$; for normal Anderson disorder,
it may be calculated using $\sigma_0\sigma_0\sigma_0=\sigma_0$. Thus, the
$\sigma_0$ term does not produce any difference for the two types of
disorder, and the Fermi energy, which corresponds to the $\sigma_0$
term, has the same renormalization for normal Anderson disorder,
$\sigma_0$ and bond disorder, $\sigma_x$. However, the $\sigma_z$ term
in the self-energy $\Sigma(E,k)$ is changed differently by Anderson
disorder and bond disorder. Because
$\sigma_0\sigma_z\sigma_0=\sigma_z$ and
$\sigma_x\sigma_z\sigma_x=-\sigma_z$, the topological mass, which
corresponds to the $\sigma_z$ term, is renormalized along opposite
directions by Anderson disorder $\sigma_0$ and bond disorder
$\sigma_x$. This can be seen clearly in Eq. (9). Hence, TAI
phenomenon in a $\sigma_x$ disordered system may manifest different
features from that in a $\sigma_0$ disordered system.

Although the above formula are derived in the HgTe/CdTe quantum wells, they should be also applicable to the Haldane model since the low energy effective Hamiltonian for the two models have the same representation as BHZ model.~\cite{6}
In addition, it should point out that the disorder-induced mass inversion always corresponds to a topological phase transition in HgTe/CdTe quantum wells,
however it is not true for the Haldane model if the topological mass only change its sign at K or K$^*$ point.
That is because the topological property of the Haldane model is determined by the relative sign of the topologically effective mass
at the two Dirac points K and K$^*$. Namely, if the topological masses have opposite sign at K and K$^*$ points, the system is topologically non-trivial. Otherwise, the system is topologically trivial. In the following, we will see clearly that the disorder, regardless of Anderson disorder or bond disorder, has the same effects on the two models.

From the models in Eqs. (1) and (2), it is easy to describe a nanoribbon geometry as shown in
Fig. 1. Here, only the
nanoribbon with a zigzag edge is studied for the Haldane model. Using the
Landauer-B\"{u}ttiker formula, the linear conductance at zero
temperature and low bias voltage can be represented
as:~\cite{Jiang,Pareek,SDatta}
\begin{eqnarray}
G_{LR} & =&\frac{e^2}{h}T=\frac{e^2}{h}Tr{[\Gamma_LG^r\Gamma_RG^a]},
\end{eqnarray}
where $T=Tr{[\Gamma_LG^r\Gamma_RG^a]}$ is the transmission
coefficient from the left lead (source) to the right lead (drain),
$\Gamma_{L/R}=i(\Sigma^r_{L/R}-\Sigma^a_{L/R})$ with
$\Sigma^{r/a}_{L/R}$ being the retarded/advanced self energy,
respectively. To present a more concrete picture, we will provide some numerical calculations for the
two models and follow with discussions on the implications in section III.

\section{Numerical Results and Discussions}
\label{sec:discussions}
\subsection{Band structures for two models}
In this section, the conductance, conductance fluctuations and
localization length are studied numerically. The size of the central
region is denoted by integers N and W, which represent the length and
width respectively. For example, in the schematic diagram shown in Fig. 1(a), the
length of the central region (red) is given by $L_x=N\times a$
with $N=9$, and the width by $L_y=W\times a$ with $W=5$. In Fig. 1(b), the length of the central region (red) is given by
$L_x=N\times \sqrt{3} a$ with $N=5$, and the width by $L_y=W\times \sqrt{3}a$ with $W=3$. Here, $a$ is
the square lattice constant $a=5$nm for the HgTe/CdTe model and
represents the nearest neighboring distance of $a=0.142$nm for the Haldane model.
\begin{figure}[!ht]
\includegraphics[width=1.0\columnwidth, viewport=33 220 903 636, clip]{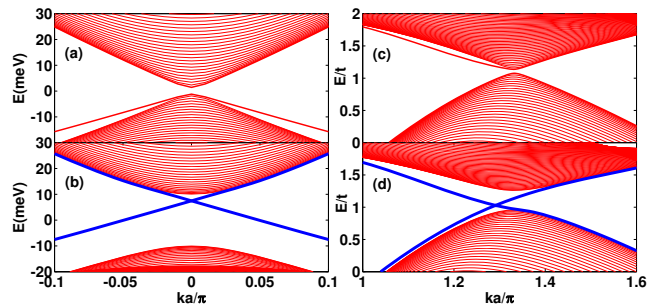}
\caption{ (Color online) The band structures for the HgTe/CdTe and
Haldane models are shown in (a), (b) and (c), (d). (a) and (b) are plotted with different
  topological masses of $M=1$meV and $M=-10$meV; (c) and (d) are plotted with
 different staggered terms of $\Delta=-1.77t$ and $\Delta=-1.60t$.
The red (blue) lines represent the bulk states (edge states).}
\end{figure}

We have plotted the band structures of the two models in Fig. 2. Figs.
2(a) and 2(b) are plotted for the HgTe/CdTe model with parameters
$M=1$meV 2(a) and $M=-10$meV 2(b); Figs. 2(c) and 2(d) are for the
Haldane model with $\Delta=-1.6t$ 2(c), and $\Delta=-1.77t$ 2(d). In
the numerical calculations, other sample-specific parameters are
fixed to be $A=364.5$meV$\cdot$nm, $B=-686$meV$\cdot$nm$^2$, $C=0$,
$D=-512$meV$\cdot$nm$^2$ for the HgTe/CdTe model. For the Haldane model, $t$ is
set as the energy unit ($t=1$), and other parameters are set with values of
$t_1=-t$, $t_2=-0.5t$, $\phi=-0.235\pi$. It can be clearly seen
that for the topologically nontrivial phase of two models, gapless states
traverse the bulk gap in Figs. 2(b) and 2(d). When
changing the topological mass from the topologically nontrivial phase to the topologically
trivial phase, gapless states disappear from the bulk gap
for 2(a) and 2(c). If the stripe geometry in Figs. 1(a) and 1(b) has
a periodic boundary in the $y$ direction, namely transforming the system into a cylindrical
geometry, the gapless edge states disappear in 2(b) and 2(d), but no
visible changes can be observed in 2(a) and 2(c). For simplicity, the band structures for the
cylindrical geometry are not shown here. This means that the two
systems are both 2D topologically trivial insulators for the parameters in
Figs. 2(a) and 2(c) but are topologically nontrivial insulators for the parameters in Figs. 2(b) and 2(d).

\subsection{The conductance and its fluctuations}
In Figs.(3) and (4), we investigate the conductance and its fluctuations
versus strength of the two types of disorder in the models.
In all of these calculations data are averaged for 500 random
configurations of disorder.
Note that for Anderson disorder, the results show the
same physics as that described in previous papers~\cite{Shen,Jiang,Beenakker} for the HgTe/CdTe model [see
Figs. 3(a) and 3(c)]. The conductance and its fluctuations are specifically studied for the Haldane
model with Anderson disorder [see Figs. 4(a) and 4(c)] and the same
quantum conductance plateau and zero conductance fluctuation can
be seen within a definite range of Anderson disorder. Thus, for Anderson
disorder, we find the following properties: First, the conductance
decays and its fluctuations initially increases gradually when increasing
strength of Anderson disorder. Second, at
moderate Anderson disorder strength, the conductance stops deceasing
and falls to a quantum plateau ($e^2/h$, when considering only spin up or
spin down case). In addition, the conductance fluctuation reduces to zero [see
Fig. 3(c)] or a very small value [see Fig. 4(c)] over the
corresponding range of Anderson disorder strength.~\cite{Note25} Third, as the strength of Anderson disorder continues to increase,
the conductance and its fluctuations both decrease gradually to zero. The system thus finally
transforms to an Anderson insulator when the disorder is sufficiently strong.
\begin{figure}[!ht]
\includegraphics[width=1.0\columnwidth, viewport=25 8 804 630, clip]{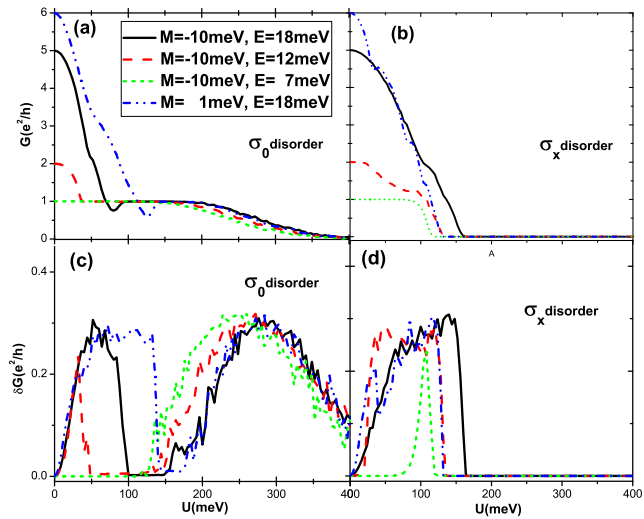}
\caption{ (Color online) The conductance $G$ (a), (b) and
conductance fluctuations $\delta G$ (c), (d) vs disorder strength U
for the HgTe/CdTe model. Different curves correspond different Fermi
energy $E$ or different topological masses. The size of the central region
is set at $L_x=200a$, $L_y=80a$. Other parameters (A, B, C and D) are
given at the beginning of Sec.~\ref{sec:discussions}. Here, $\sigma_0$ and $\sigma_x$
disorders correspond to Anderson disorder and bond disorder
respectively.}
\end{figure}

In general, the quantum conductance
plateau and zero conductance fluctuations always imply a new phase or a novel phenomenon.
From the results about the quantum conductance, TAI was initially found three years ago.~\cite{Shen}
As mentioned in Introduction, Groth {\sl et al.}~\cite{Beenakker} showed how Anderson disorder induce a phase transition.
Namely, Anderson disorder will add a negative correction to the topological mass. When the topological mass
changes its sign at strong Anderson disorder strength, a
phase transition is triggered from the topologically trivial phase to the topologically nontrivial phase.

However, due to the extensive existence of bond disorder in real materials,
it is necessary and important to study what happens when considering bond disorder. In Figs. 3(b), 3(d), 4(b), and 4(d), the
conductance and its fluctuations are investigated for the two models
with bond disorder. Contrary to the case of Anderson disorder,
TAI is not observed in the two models when changing the strength of
bond disorder. For example, for the HgTe/CdTe model in Figs. 3(b) and
3(d), a series of topological masses $M$ and Fermi energies $E$
are chosen. In each case, the conductance gradually falls to zero and the quantum
plateau shown in Fig. 3(a) disappears completely. Moreover, the
conductance fluctuations in Fig. 3(d) only shows a peak and then falls to zero finally because of the Anderson
localization. This description is also qualitatively true for the Haldane
model in Figs. 4(b) and 4(d). Therefore, it can be concluded that bond disorder cannot induce a phase transition to TAI
for the two models as discussed above in Sec. II.

\begin{figure}[!ht]
\includegraphics[width=1.0\columnwidth, viewport=70 60 760 595, clip]{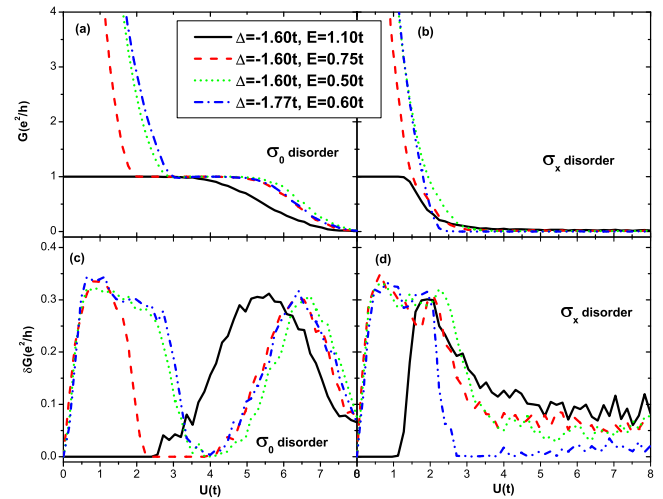}
\caption{ (Color online) The conductance $G$ (a), (b) and
conductance fluctuations $\delta G$ (c), (d) vs disorder strength U
for Haldane model. Different curves correspond to different Fermi
energies $E$ or staggered terms $\Delta$. The size of the central region is
set to be $L_x=200\times \sqrt{3} a$, $L_y=92\times 3a$. Other parameters
($t_1$, $t_2$, and $\phi$) are given at the beginning of Sec.~\ref{sec:discussions}.
Here, $\sigma_0$ and $\sigma_x$ correspond to Anderson
disorder and bond disorder respectively.}
\end{figure}

\subsection{The conductance phase diagrams for two types of disorder}
To see this more clearly, the conductance phase diagrams are compared for
Anderson disorder and bond disorder in the HgTe/CdTe model. The results,
in Figs. 5(a), 5(c), 6(a), and 6(c), are shown for normal
Anderson disorder; Figs. 5(b), 5(d), 6(b), and 6(d) correspond to
bond disorder. It can be seen from Figs. 6(a), and 6(c) that at
moderate Anderson disorder and Fermi energy, a clear TAI phase (green region) is present and that the TAI phase in Fig. 6(c) must correspond to a negative
renormalized topological mass $\overline{M}$ (blue region). The Anderson disorder renormalizes the topological mass $M$ along the negative direction and Fermi energy along the positive direction and therefore induced a phase transition to TAI. The results in Figs. 5(a), 5(c), 6(a), and 6(c) are similar to those described in previous
papers.~\cite{Shen,Jiang,Beenakker} In summary, normal Anderson
disorder can localize the bulk states. For a original system of
a topological nontrivial phase [Fig. 5(a)], the edge state is more robust against
Anderson disorder than are bulk states and thus leads to a quantum
conductance region (green region). For a original system with a topologically
trivial phase [Fig. 6(a)], a phase transition from a topologically
trivial insulator to a topological insulator can occur as the strength of Anderson disorder increases, and thus result in a
TAI quantum conductance region (green region). For Anderson
disorder, the topological mass is always renormalized by adding a
negative correction. A more detailed interpretation can be found in
several previous papers.~\cite{Shen,Jiang,Beenakker}

\begin{figure}[!ht]
\includegraphics[width=1.0\columnwidth, viewport=75 55 685 490, clip]{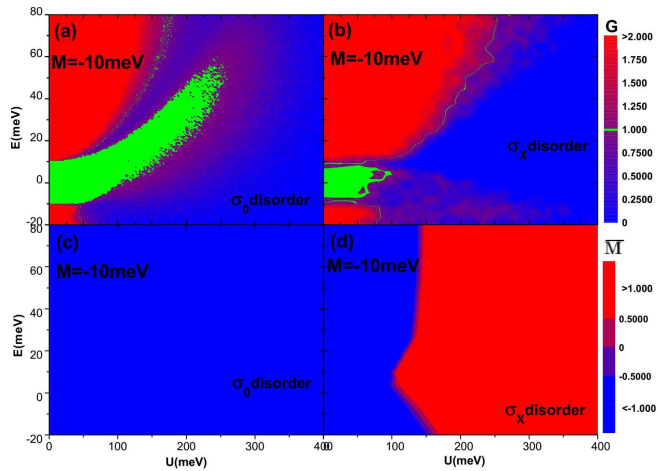}
\caption{ (Color online) The conductance $G$ (a), (b) and
topological mass $\overline{M}$ (c), (d) vs disorder strength U for the
HgTe/CdTe model with $M=-10$meV. The size of the central region is set to be
$L_x=200a$, $L_y=80a$. Other parameters (A, B, C, and D) are given at
the beginning of Sec.~\ref{sec:discussions}. Here, $\sigma_0$ and $\sigma_x$
correspond to Anderson disorder and bond disorder
respectively.}
\end{figure}

In the following, we are more concerned about the influence of bond disorder on
this phenomenon. It can be seen from Fig. 5 that at $M=-10$meV, the
quantum conductance region with a value of $e^2/h$ (green  region) above
the band gap in Fig. 5(a) disappears completely as shown in Fig. 5(b).
Namely, the quantum conductance region where the value of the conductance is
$e^2/h$ (green region) decays to a very small region, and only exists
within the band gap and at a small strength of bond disorder [Fig.
5(b)]. The renormalized mass $\overline{M}$, which remains
negative for the whole phase diagram of Fig. 5(c) (blue  region),
alters its sign between the range of $100$ meV $\lesssim U
\lesssim 150$meV in Fig. 5(d). Thus, there are very different
effects on the mass $M$ for the two types of disorder.

It may be surprising that a larger phase region of bulk states (red
 region) appears in Fig. 5(b) than in Fig. 5(a).
This implies a greater difficulty in localizing the bulk states for
the case of bond disorder compared to the case of Anderson
disorder. This result is understandable if the different
effects of Anderson disorder and bond disorder are considered.
Anderson disorder renormalizes the topological mass $M$ along
positive direction. That is, the effective band gap
became larger and thus bulk states are shifted upward for the Anderson
disorder case. On the contrary, the opposite effect appeared for the bond disorder. Because states in the center of the bulk band are
more difficult to localize than states near the edge of bulk
band,~\cite{Thouless} a larger phase region of bulk states is
observed in Fig. 5(b) for bond disorder than that in Fig. 5(a) for
Anderson disorder. Note that the conclusions above are based on the
same renormalization of the chemical potential $\overline{\mu}$ for both
disorder types as given in Eq. (10).

\begin{figure}[!ht]
\includegraphics[width=1.0\columnwidth, viewport=75 60 680 490, clip]{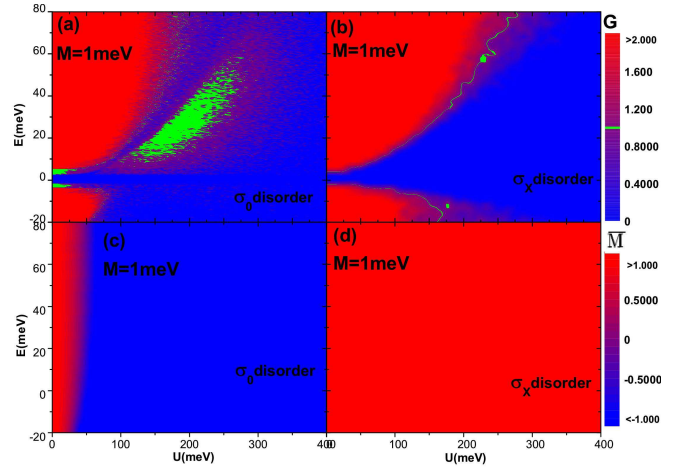}
\caption{ (Color online) The conductance $G$, (a), (b) and
renormalized topological mass $\overline{M}$ (c), (d) vs disorder
strength U and Fermi level $E$ for the HgTe/CdTe model with $M=1$meV.
The size of the central region is set to be $L_x=200a$, $L_y=80a$. Other
parameters (A, B, C, and D) are given at the beginning of Sec.~\ref{sec:discussions}.
Here, $\sigma_0$ and $\sigma_x$ correspond to the Anderson
disorder and bond disorder respectively.}
\end{figure}

When the system turns to a topologically trivial phase with
$M=1$meV as in Fig. 6, there is no indication of the TAI phenomenon
in the conductance phase diagram, Fig. 6(b). That is, the
conductance region with the value of $e^2/h$ disappears completely for
bond disorder. Meanwhile, in the phase diagram with a topological
mass $M=1$meV as shown in Fig. 6(d), the topological phase (blue region)
gives way to the topologically trivial phase (red region) and the
whole diagram is governed by positive renormalized topological mass
(red region). Therefore, a clear conclusion can be drawn from these
numerical results that normal Anderson disorder gives rise to the
TAI phenomenon and bond disorder destroys it. In other words, whether the TAI phenomenon can be observed depends on a
competition between Anderson disorder and bond disorder in a system because Anderson disorder and bond disorder renormalize the
topological mass along different directions and thus affect TAI phenomena differently.

\begin{figure}[!ht]
\includegraphics[width=1.0\columnwidth, viewport=75 60 670 490, clip]{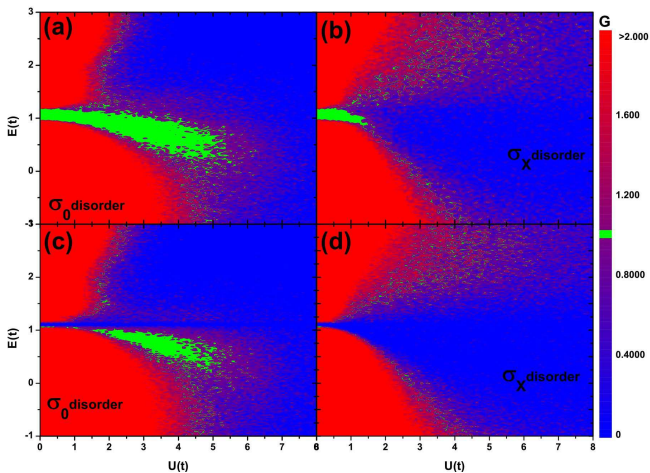}
\caption{ (Color online) The conductance $G$ vs. disorder strength U
and Fermi level $E$ for the Haldane model. (a) and (b) are plotted with
$\Delta=-1.6t$ in a topologically nontrivial region; (c) and (d) with
$\Delta=-1.77t$ in a topologically trivial region. The size of the central
region is set to be $L_x=200\times \sqrt{3} a$, $L_y=92\times 3a$. Other
parameters are chosen with $\phi=-0.235\pi$, $t_2=-0.5t$. Here,
$\sigma_0$ and $\sigma_x$ correspond to the Anderson
disorder and bond disorder respectively.}
\end{figure}

The above conclusions  are also applicable to the Haldane model. When considering Anderson disorder
in Fig. 7(a) with the topologically nontrivial phase and 7(c) with the topologically trivial phase, the conductance
phase diagrams show a clear region with quantize conductance $G_0$ (green region), which is the hallmark of TAI in Fig. 7(c).
However, for bond disorder, the region of quantized conductance becomes indistinguishable in Fig. 7(b)
for the topologically nontrivial phase and in 7(d) for the topologically trivial phase. This indicates that Anderson
disorder can drive a phase transition to TAI but that bond disorder can not. Here, we do not
give a topological mass diagram similar to Fig. 5(c) or 5(d) for the HgTe/CdTe model because the Haldane
model is based on the honeycomb lattice which means that the topological properties of this model are determined by
the signs of the effective masses at the two Dirac points, as introduced in Sec.~\ref{sec:models}. When the signs of the effective masses
at two Dirac points are opposite, the band at one Dirac point would be inverted from that at another Dirac point, and
the system is topologically non-trivial. Conversely, when the sign of the effective mass at two Dirac points is the same,
 the system corresponds to a topologically trivial phase. In addition, the parameter $\beta$ in Eq. (5) has different expressions
at the two the Dirac points. Consequently, it is difficult to present clearly the variation of the topological properties as a function of
the disorder strength using a simple phase diagram of the topological mass.

\subsection{The localization lengths for two types of disorder}
In order to highlight the dependence of TAI on the type of disorder,
we also plot the localization length~\cite{Beenakker,MacKinnon,Kramer,Pichard}
$\lambda /L_y$ of a 2D ribbon with width $L_y = W\times a$ for
the square lattice [Figs. 8(a) and 8(c)] and $L_y= W \times 3a$
for the honeycomb lattice [Figs. 8(b) and 8(d)]. Note that the localization length is obtained from the relationship~\cite{Beenakker}
$\lambda\equiv 2$lim$_{L_x\rightarrow \infty}L_x\langle $ln$(G/G_0)\rangle^{-1}$ by increasing the length $L_x$ of the system at fixed width $L_y$,
namely, $L_x\gg L_y$. In addition, another calculating method, transfer-matrix method,~\cite{MacKinnon,Kramer,Pichard} is also adopted
to guarantee the correctness of the localization length. Here we have
considered both cylindrical geometry and ribbon geometry.

For the HgTe/CdTe model, the localization length $\lambda
/L_y$ shown in in Fig. 8(a) first deceases when increasing disorder strength for Anderson disorder.
With further increasing the strength of Anderson disorder,
a peak appears for the localization length $\lambda
/L_y$ and then decays gradually for both geometries, which can be seen
clearly in Fig. 8(a). The difference in
localization length $\lambda /L_y$ between the two geometries is that
peaks have distinct heights and are located at the different strengths of
Anderson disorder. It can be deduced that because of existing edge
states, a special metal phase emerges for the ribbon geometry near to the position of
the peak. However, when considering bond disorder,
the peak of the localization length $\lambda /L_y$ vanishes
completely as shown in Fig. 8(c). This provide a further evidence that bond disorder does not induce
a phase transition to TAI. As is shown in Fig. 8(b) and 8(d) for the Haldane model,
there is no essential difference with the results shown in Figs. 8(a) and 8(c) for the HgTe model.

\begin{figure}[!ht]
\includegraphics[width=1.0\columnwidth, viewport=66 55 790 604, clip]{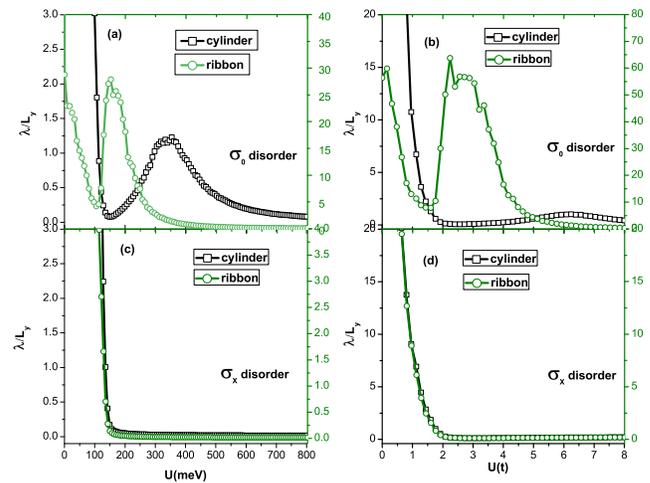}
\caption{ (Color online) Localization length vs disorder strength U for
the HgTe/CdTe model (a), (c) and for the Haldane model (b), (d). (a), (b) and
(c), (d) correspond to $\sigma_0$ and $\sigma_x$ disorder
respectively. In the calculations, the Fermi energy $E$ is set to be
$E=25$meV, the topological mass is $M=-10$meV and the width of the ribbon is
$L_y=50a$ for the HgTe/CdTe model. The Fermi energy is $E=0.80t$, the
staggered term is $\Delta=-1.60t$ and the width of the ribbon is $L_y=40\times 3
a$ for the Haldane model. The black line and green line are plotted for the
ribbon and cylindrical geometry respectively. Here, $\sigma_0$ and
$\sigma_x$ correspond to the Anderson disorder and bond
disorder respectively.}
\end{figure}

\section{Conclusions}
\label{sec:conclusions}
In summary, we have studied the influence of bond disorder (called
$\sigma_x$ disorder) on two models. This
type of disorder can originate in the deformation of the lattice,
mismatch between two lattices or some chemical effects and thus
exists extensively in many real materials. Unlike normal Anderson
disorder, bond disorder does not induce a
phase transition to a topological Anderson insulator(TAI). This is analytically
verified by using Born approximation theory. The conductance
and its fluctuations are then calculated, and the conductance plateau, which is the hallmark of TAI, is very good for Anderson disorder but
becomes barely distinguishable for bond disorder. In addition, phase
diagrams are compared for two models with two types of disorder,
Anderson disorder $\sigma_0$ and bond disorder $\sigma_x$. For
Anderson disorder, the TAI phase can be seen clearly in the phase diagrams. However, the TAI phase disappears for bond disorder.

Because of a effective characterization in the metal-insulator phase transition,
the localization length of the wave function is studied for the two models with Anderson and bond disorders.
For the two models with normal Anderson disorder, the localization length shows a peak at moderate disorder region, which exactly
corresponds to TAI phase. However, the wave functions are localized quickly and no peak of localization length can be seen for the system
with bond disorder. That means that Anderson disorder and bond disorder can play different roles in the topological phase transition.

To sum up, bond disorder can prohibit the system undergoing a phase
transition to TAI, contrary to what may be seen in a Anderson
disordered insulator.~\cite{Note30} These general conclusions in this paper are not
restricted to two-dimensional systems and further work in a three-dimensional systems,
e.g. $Bi_2Se_3$ model,~\cite{4} will be needed. If researchers intend to change the property of a
topological insulator by addition of impurities, eg. shifting the energy level
from the bulk states to the bulk gap, it will be necessary to guarantee that the impurities are mainly
of the Anderson disorder type but keeps away from the bond disorder type. This
is a key point for the doping process in preparation of topological
insulators.

\section*{ACKNOWLEDGMENTS}
We are grateful to Dongwei Xu and Yanyang Zhang for their helpful
discussions. Juntao Song is supported by NSFC under Grant No.
11047131 and RFDPHE-China under Grant No. 20101303120005. Hua Jiang
is supported by China Post-doctroal Science Foundation under Grant
No. 20100480147 and No. 201104030. Qing-feng Sun and X.C. Xie are supported by NSFC under
Grant No. 10821403, No. 10974236,  and China-973 program.

\end{document}